\documentclass[aps,twocolumn, prl, footinbib]{revtex4-1}
\usepackage{amsmath,amssymb} \usepackage{colordvi} \usepackage{color}
\usepackage{graphicx}
\usepackage{dcolumn}
\usepackage{bm}
\usepackage{romannum}
\usepackage{caption}
\usepackage{subcaption}
\usepackage{cleveref}
\usepackage{csquotes}
\usepackage{float}
\usepackage{color,soul}

\begin{document}
 \draft
 \title{One- and Two-Particle Problem with Correlated Disorder Potential}
 \author{Guangcun Liu}
 \address{Laboratory of Quantum Engineering and Quantum Metrology, School of Physics and Astronomy, Sun Yat-Sen University (Zhuhai Campus), Zhuhai 519082, China}
\date{\today}

\begin{abstract}
Motivated by the recent experimental and theoretical progresses in the exploration of the effect of disorder in interacting system, we examine the effect of two types of correlated disorder, the quasi-periodic potential and speckle disorder potential, on one- and two-particle problem with exact diagonalization (ED) method. We give the phase diagram for single particle in the presence of quasi-periodic potential and also analyse the effect of strong interaction on the phase diagram for ground state in two dimensions. For the speckle disorder potential case, we examine both the effect of correlation length and disorder strength on single particle ground state energy and two-particle binding energy. The transport property for different interaction strength under speckle disorder potential is also calculated and discussed at last. 
\end{abstract}

\maketitle
\section{\label{sec:level1}I. introduction}
Disorder has been an important topic for a long time because of its essential role in our understanding of many different phenomena such as conductor-insulator transition, mesoscopic phenomena and so on~\cite{50anderson,conductorinsulator,
mesoscopic1,mesoscopic2}. For disordered system without interaction, Anderson localization \cite{anderson} has been one of the most important topic which has been studied for a long time and many interesting results have been obtained. P. W. Anderson had shown that disorder could result in an exponentially localized wave function which leads to the zero conductivity. Later, the scaling theory was proposed which showed that for one and two-dimensional case, the wave function of the particle is localized for any non-zero random disorder strength while for three-dimensional case, it requires a large enough disorder strength to localize the wave function~\cite{scaling}. 

Many experiments have been designed to explore Anderson localization and it has been observed in classic waves including light~\cite{lightlocalization1,lightlocalization2}, microwave~\cite{microwavelocalization} and sound~\cite{soundlocalization}. In ultracold atoms, direct observation of Anderson localization has been realized since the wave function can be observed and its evolution can be monitored directly. Anderson localization has been observed in experiments conducted with ultracold atoms in both one dimension \cite{1dal1,1dal2} and three dimensions \cite{3dal1,3dal2}.   

It should be emphasised that the on-site disorder potential used in Anderson's work \cite{anderson} is generated randomly from a uniform distribution of width $W$ which means there is no correlation from site to site. On the other hand, different with the widely accepted theorem that any non-zero random disorder in one dimension would result in exponentially localized state, correlations on the disorder distribution could bring surprising effect such as delocalization effect which leads to the improvement of transport property or localization-delocalization transition \cite{diagonaloffdiagonal,RDM1,RDM2,singlespeckle}. This is also one of the motivations for us to study correlated disorder potential.

Many experiments in ultracold atoms have been conducted in order to shine light on the effect of disorder in superfluid-insulator transition, localization-delocalization transition. Some of these experiments are directly related to the confirmation of the existence of many-body localization state \cite{1d interaction AAH,isingexperiment,quasi2d interacting AAH,2d random disorder experiment,2d quasi experiment}. Some of the experiments are connected with the measurement of transport property \cite{conduction,RC,bosons,fermions}. Motivated by these experiments, we mainly consider two types of correlated disorder potential which are widely used in these experiments. The first is the quasi-periodic potential which is usually related to Harper model \cite{harper,aubry andre}, the second is the speckle disorder potential \cite{specklepattern,simulationspeckle}. We examine the effect of correlated disorder on one- and two-particle problem in two dimensions with exact diagonalization method \cite{computationalmanyparticle}. In our calculation, the lattice size is usually $64*64$ and we usually average 50 disorder realizations for one-particle case and 10 disorder realizations for two-particle case. 
The comparison with one-dimensional results \cite{tworandom,imry,oppen,twoharper,ueffect,andersonlocalizationofpair,onedrudeweight,frahm,
weinmann,bosonspeckle,1d boson} could be helpful to the understanding of the role of dimensionality as well as correlated disorder in the interacting system. We also wish the results from one- and two-particle disordered system could be helpful for the understanding of recent disordered experimental results \cite{2d quasi experiment,fermions}. In these experiments, temperature is usually very low and as a result,  we focus on ground state of single and two-particle case in two dimensions. As shown in the following, we will show that these ground state results could be enlightening for our understanding of these experimental results. 

This paper is organized as follows. We first consider the single and two-particle problem in quasi-periodic disorder potential in Sec. II. Through comparing the single and two-particle results, we show the effect of strong interaction in this case. Then, in Sec. III, with similar strategy, we consider single and two-particle problem in speckle disorder potential. Through comparing the single and two-particle results, we also show the effect of strong interaction in speckle disorder potential case. We also examine the transport property in this case. Finally, we provide a summary in Sec. IV. 
\section{\label{sec:level1} II. Quasi-periodic disorder potential}
One-dimensional Harper model \cite{harper,aubry andre} is widely investigated because of its phase transition point between delocalized phase and localized phase when quasi-periodic potential $\Delta$ equals $2$ (in unite of the hopping term). Motivated by the recent experiment \cite{1d interaction AAH,quasi2d interacting AAH,2d quasi experiment}, we consider the two dimensional Harper model. 
The Hamiltonian for single particle is: 
\begin{equation}
\begin{aligned}
\hat{H}_{0}&=-t\sum_{\langle i,j\rangle,\sigma}(\hat{c}_{i,\sigma}^{+}\hat{c}_{j,\sigma}+h.c.)+ \\
&\sum_{i,\sigma}\Delta_{x} \cos(2\pi\beta m+\varphi) n_{i,\sigma}+
\sum_{i,\sigma}\Delta_{y} \cos(2\pi\beta n+\varphi) n_{i,\sigma}
\end{aligned}
\end{equation}
The first two terms are the hopping between nearest-neighbor sites. $\hat{c}^+_{i,\sigma}$ ($\hat{c}_{i,\sigma}$) indicates the creation (annihilation) operator for a fermion with spin $\sigma$ on lattice site $i$. $\Delta_{x}$ $(\Delta_y)$ controls the disorder strength of the potential in $x$ $(y)$ direction. $n_{i,\sigma}=\hat{c}^+_{i,\sigma}\hat{c}_{i,\sigma}$ is the number operator at site $i$ with spin $\sigma$. $(m,n)$ is the row and column index for lattice site $i$. The potential is the quasi-periodic potential with $\beta$ equals to $(\sqrt{5}-1)/2$ and $\varphi$ a uniform phase.  

As widely used in one dimension, inverse participation ratio (IPR) is introduced to study the possible localization-delocalization transition,
\begin{equation}
\alpha_{p}=\sum_{i=1}^L |\psi(i)|^4,
\end{equation}
here $\psi(i)$ is the single particle wave function, $L$ is the lattice size of the system. $\alpha_{p}$ can be used as order parameter to distinguish localized state from extended state since it is finite in the localized state and vanishes in the extended state.

\begin{figure}[H]
\centering
\includegraphics[scale=0.5]{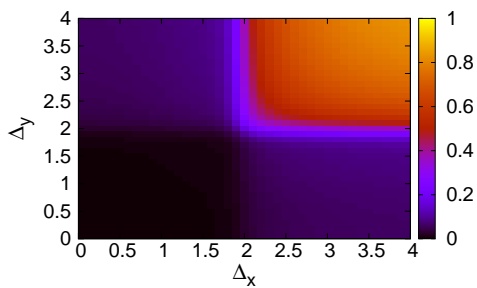}
\caption{(Color online) Inverse participation ratio for single particle two-dimensional Harper model.}
\label{phase1}
\end{figure}

The result for the single particle inverse participation ratio is shown in Fig.~\ref{phase1}. The result shows that there are three different phases which is different with one-dimensional case where only two phases (localized and delocalized phase) are present. Phase I ($\Delta_{x}<2.0$ and $\Delta_{y}<2.0$) is a delocalized phase while phase II ($\Delta_{x}>2.0$ and $\Delta_{y}>2.0$) is a localized phase. Phase III (($\Delta_{x}>2.0$ and $\Delta_{y}<2.0$) and ($\Delta_{x}<2.0$ and $\Delta_{y}>2.0$)) is an intermediate phase since it is localized in one direction and is delocalized in the other direction. 
When the disorder strength in the delocalized direction becomes larger than 2, both directions become localized and give phase II.
This two-dimensional single particle case gives a simple but standard picture to explore two-dimensional delocalization-localization transition and could be helpful to investigate more complicated cases. 

In addition, we examine the effect of strong interaction on two-dimensional Harper model, we consider two fermions (one spin up and one spin down) with large on-site attractive interaction under two-dimensional quasi-periodic potential:
\begin{equation}
\hat{U}=-|U|\sum_{i}n_{i\uparrow}n_{i\downarrow}, 
\end{equation}
the interaction strength is set to be $|U|=5t$. The Schr\"odinger equation can be written as $(E-\hat{H}_0)|\Psi\rangle=\hat{U}|\Psi\rangle$ and $\Psi$ is the two-particle wave function. 
Following the procedure of ref. \cite{andersonlocalizationofpair}
one could obtain:
\begin{equation}
-\frac{1}{|U|}\langle i,i'|\Psi\rangle=\sum_j\langle i,i'|\hat{G}_{E}|j,j\rangle\Psi(j,j).\label{sge}
\end{equation}
$\hat{G}_{E}=(E-\hat{H}_0)^{-1}$ is Green's function and $\Psi$ is the two-particle wave function. $|j,j\rangle$ means that the first and the second particle locate on the same lattice site $j$. 
On the other hand, one could obtain:
\begin{equation}
\hat{G}_{E}=\sum_{r,s}(E-\varepsilon_{r}-\varepsilon_{s})^{-1}|\psi_{r},\psi_{s}\rangle\langle\psi_{r},\psi_{s}|,
\end{equation}
$\varepsilon_r$ and $\varepsilon_s$ are the eigenvalue of the single particle state of $\psi_r$ and $\psi_s$. Projecting the Green's function on the two non-interacting particle bases, one could obtain:
\begin{equation}
\begin{aligned}
\langle i,i'|\hat{G}_{E}|j,j\rangle=&\sum_{r,s}\frac{\langle i,i'|\psi_{r},\psi_{s}\rangle\langle\psi_{r},\psi_{s}|j,j\rangle}{E-\varepsilon_{r}-\varepsilon_{s}} \\
&=\sum_{r,s}\frac{\psi_{r}(i)\psi_{s}(i')\psi_{r}^{*}(j)\psi_{s}^{*}(j)}{E-\varepsilon_{r}-\varepsilon_{s}}.
\end{aligned}\label{G}
\end{equation}
Setting $i'=i$, and put Eq. (\ref{G}) into Eq. (\ref{sge}), we have 
\begin{equation}
-\frac{1}{|U|}f(i)=\sum_{j}K_{E}(i,j)f(j), \label{SE}
\end{equation}
here $K_{E}(i,j)=\langle i,i|\hat{G}_{E}|j,j\rangle$, $f(j)=\Psi(j,j)$. 

For two-particle problem, we also use the inverse participation ratio to study the possible localization-delocalization transition for the ground state:
\begin{equation}
\alpha_{p}=\sum_{j=1}^L |f(j)|^4.
\end{equation}

For the ground state, $\alpha_p$ can be obtained with the following procedure \cite{andersonlocalizationofpair}. For two strongly attractive particles, they will form bound state and the two-particle ground state energy should be much lower than twice of single particle ground state energy ($E < 2\epsilon_1$). Then, one could iterate the value of E and exact diagonalize the kernel until the lowest eigenvalue of the kernel equals $1/U$, then one find the ground state energy E of the two attractive energy and accordingly the eigenvector $f(j)$, then one could obtain the result of $\alpha_p$. 

\begin{figure}[H]
\centering
\includegraphics[scale=0.5]{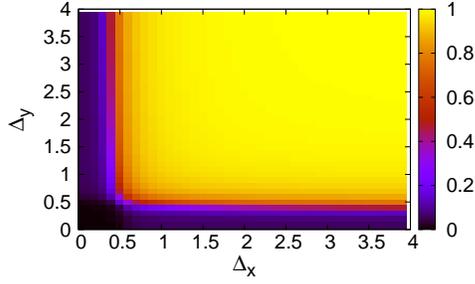}
\caption{(Color online) Inverse participation ratio for two strongly attractive particles. }
\label{phasewithinteraction}
\end{figure}

The result is shown in Fig.~\ref{phasewithinteraction}. Compared with single particle case, there are still three phases but the phase transition boundary is reduced to smaller disorder strength. This can be explained following the large $\hat{U}$ expansion procedure \cite{andersonlocalizationofpair}. In two dimensions, the expansion of the kernal becomes
\begin{equation}
\begin{aligned}
&\langle i,i|\hat{G}_E|j,j\rangle=\langle (m,n)(m,n)|\hat{G}_{E}|(k,l)(k,l)\rangle \\
&=\delta_{m,k}\delta_{n,l}\left[\frac{1}{E}+\frac{2V(i)}{E^2}+\frac{4V^{2}(i)+8}{E^3}\right]+\frac{2}{E^3}\delta_{m+1,k}\delta_{n,l} \\
&+\frac{2}{E^3}\delta_{m-1,k}\delta_{n,l}+\frac{2}{E^3}\delta_{m,k}\delta_{n+1,l}+\frac{2}{E^3}\delta_{m,k}\delta_{n-1,l}, 
\end{aligned}
\end{equation}
here $V(i)$ is the disorder potential on lattice $i$, $(m,n)$ is the row and column index for lattice $i$. Then, Eq. \eqref{SE} becomes:

\begin{equation}
\begin{aligned}
&\left[-\frac{E^2}{|U|}-E\right]f((m,n),(m,n))= \\
&\frac{2}{E}f((m\pm 1,n),(m\pm 1,n))+\frac{2}{E}f((m,n\pm 1),(m,n\pm 1))+  \\
&\left[2V(i)+\frac{4V^2(i)+8}{E}\right]f((m,n),(m,n)).\label{eff}
\end{aligned}
\end{equation}
Compare Eq. \eqref{eff} with non interacting Harper model,
one could obtain: $V_{eff}=2V(i)+[4V^2(i)+8]/E$ and $t_{eff}=-2/E$. When interaction is much larger than the hopping term and disorder strength, $E$ is approximately equal to $-|U|$. When $V_{c}$ is small as is the case when $|U|$ is large, the second order of $V(i)$ in $V_{eff}$ can be ignored. Then, Eq. \eqref{eff} becomes the effective single particle two-dimensional Harper model with the effective critical disorder strength $\Delta_{c}\approx 2/|U|$. This is in agreement with the result of inverse participation ratio calculation. 

It should be pointed out that the many-body localization experiment with two-dimensional quasi-periodic disorder potential has been conducted \cite{2d quasi experiment}. The $\Delta$ which controls the disorder strength is the same for $x$ and $y$ direction in the experiment, two quasi-periodic potentials along $x$ and $y$ directions are created and form a two-dimensional quasi-periodic potential. Through monitoring the time evolution of the particle number imbalance between even and odd stripes for different disorder strength $\Delta$, three different phases are observed (a fast thermalization regime corresponds to the delocalized phase, a negligible relaxation regime corresponds to the MBL phase and an intermediate phase with slow relaxation). These experimental results can be compared with the phase diagram for two interacting particles we obtain as shown in Fig. \ref{phasewithinteraction}. For the diagonal of Fig. \ref{phasewithinteraction} where $\Delta_x=\Delta_y$, the simulation result does not show an intermediate regime but a sharp phase transition from delocalized phase to localized phase when quasi-periodic disorder strength increases which is different from the experimental result. The intermediate phase revealed in the experiment still lacks understanding, but the difference from simulation result indicates that this intermediate phase is possible a result of the interaction effect in a real many-body system and thus can not be described by the ground state of interacting few-particle system.   

\begin{figure}[H]
\centering
\includegraphics[scale=0.4]{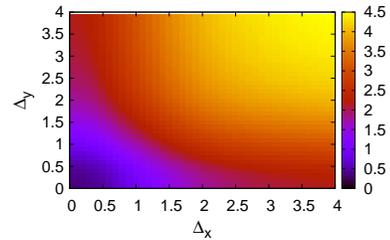}
\caption{(Color online) Binding energy for two strongly attractive fermions under two-dimensional quasi-periodic potential.}
\label{bindingenergyof2DAAH}
\end{figure}

In addition, we also calculate the binding energy for the two-particle ground state. The definition of binding energy is $E_{b}=2E_{0}-E$, here $E_{0}$ is the ground state energy of single fermion and $E$ the ground state energy of two interacting fermions. The result in Fig.~\ref{bindingenergyof2DAAH} shows that for two strongly attractive fermions in the two-dimensional Harper model, the binding energy increases with increasing disorder strength. This suggests disorder would not destroy the pair, on the contrary, disorder is helpful for the pair to be preserved. 

\section{\label{sec:level1}III. Speckle disorder potential}
\begin{figure}[h]
\centering
\begin{subfigure}[b]{0.23\textwidth}
\includegraphics[width=4.5cm]{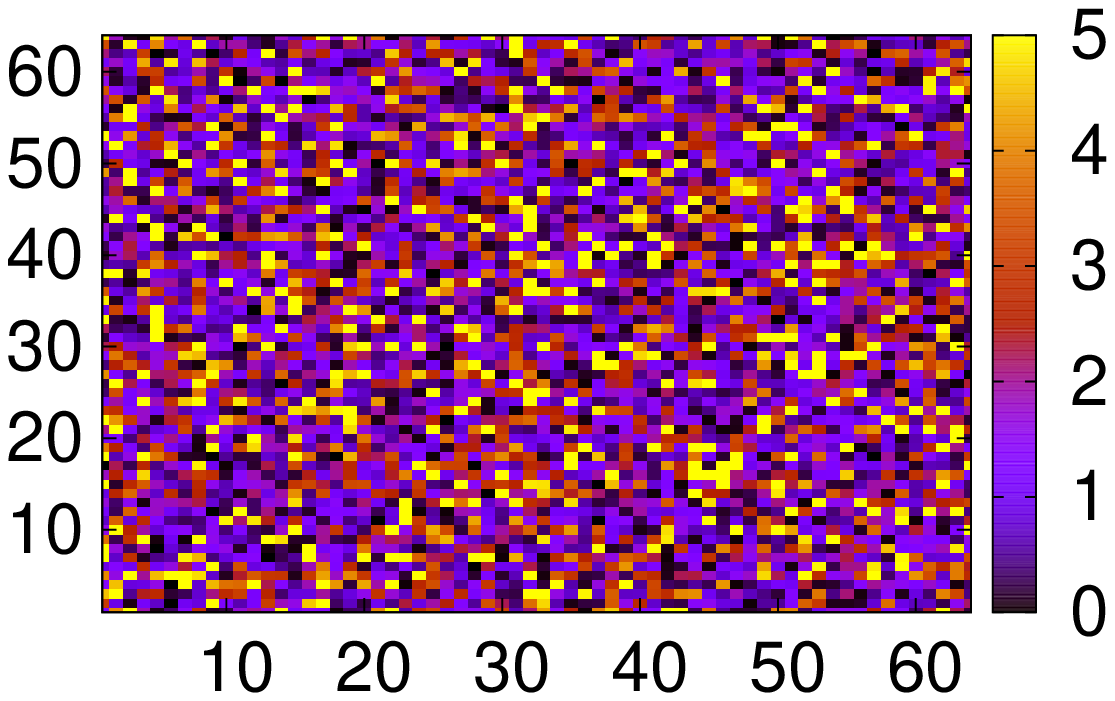}
\caption{}
\end{subfigure}
\begin{subfigure}[b]{0.23\textwidth}
\includegraphics[width=4.5cm]{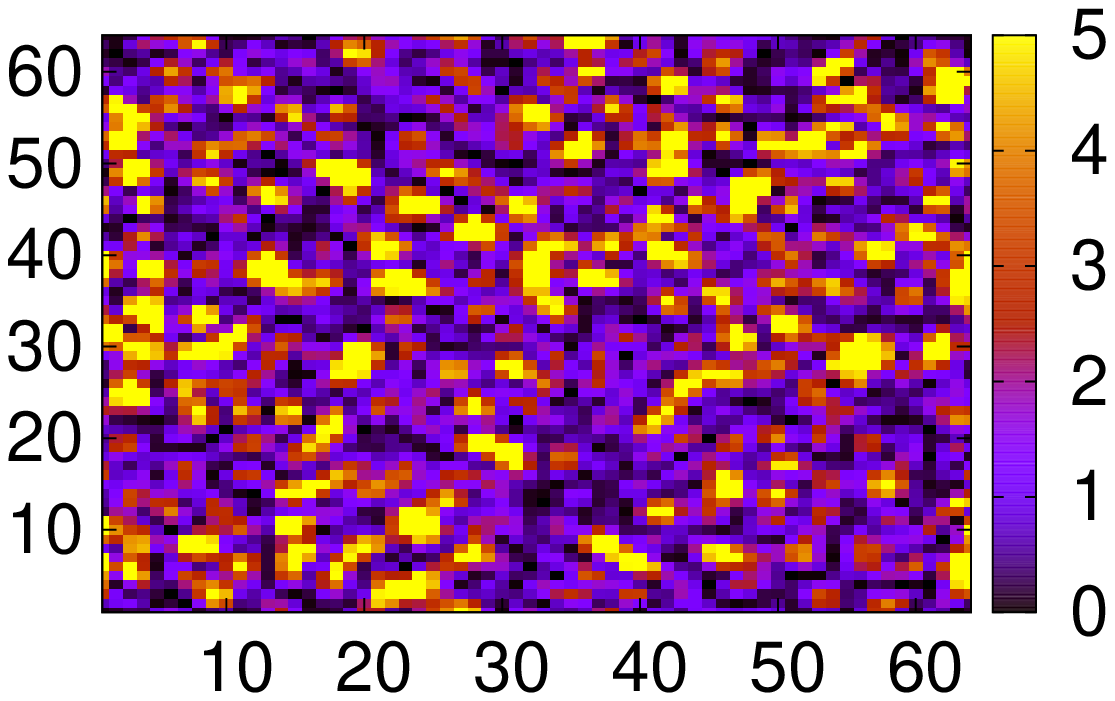}
\caption{}
\end{subfigure}
\caption{(Color online) Examples of speckle disorder potential distribution for correlation length (a) $l_c=1$ (in unit of lattice size hereafter) and (b) $l_c=5$ on the two-dimensional lattice with lattice size $64*64$. Average speckle disorder potential $\bar{V}=2.0$ (in unit of hopping term hereafter).}
\label{lcv}
\end{figure}

There are  several statistical functions that distinguish speckle disorder from random disorder such as the exponential probability distribution function and the spatial autocorrelation function of the speckle disorder \cite{specklepattern,simulationspeckle}. The correlation length $l_{c}$ and average disorder strength $\bar{V}$ are the two main quantities to characterize speckle disorder potential as shown in Fig. \ref{lcv} as illustrations. We examine the effect of both correlation length and average disorder strength on single particle ground state energy and two-particle binding energy. 

The correlation energy is defined as $E_{c}=\hbar^2/ml_{c}^2$ (we set $\hbar= m=1$ hereafter).
Then, two regions can be separated: one is the classical region ($\bar{V}\gg E_c$) where both the average disorder strength and correlation length could be large. The opposite is the quantum region ($\bar{V}\ll E_c$) \cite{quantumclassicregion}.

\begin{figure}[H]
\centering
\begin{subfigure}[b]{0.23\textwidth}
\includegraphics[width=4.0cm]{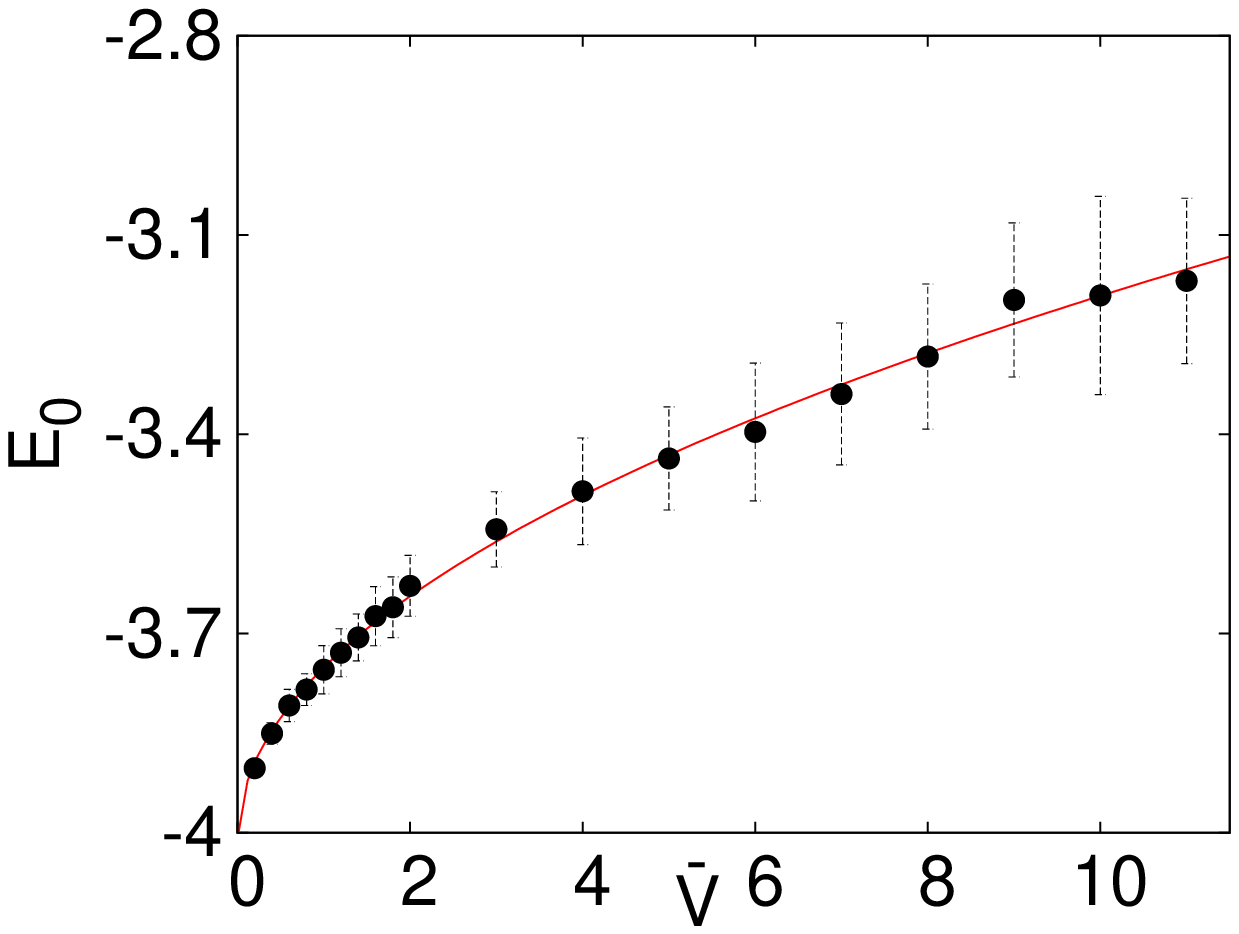}
\caption{}
\end{subfigure}
\begin{subfigure}[b]{0.23\textwidth}
\includegraphics[width=4.0cm]{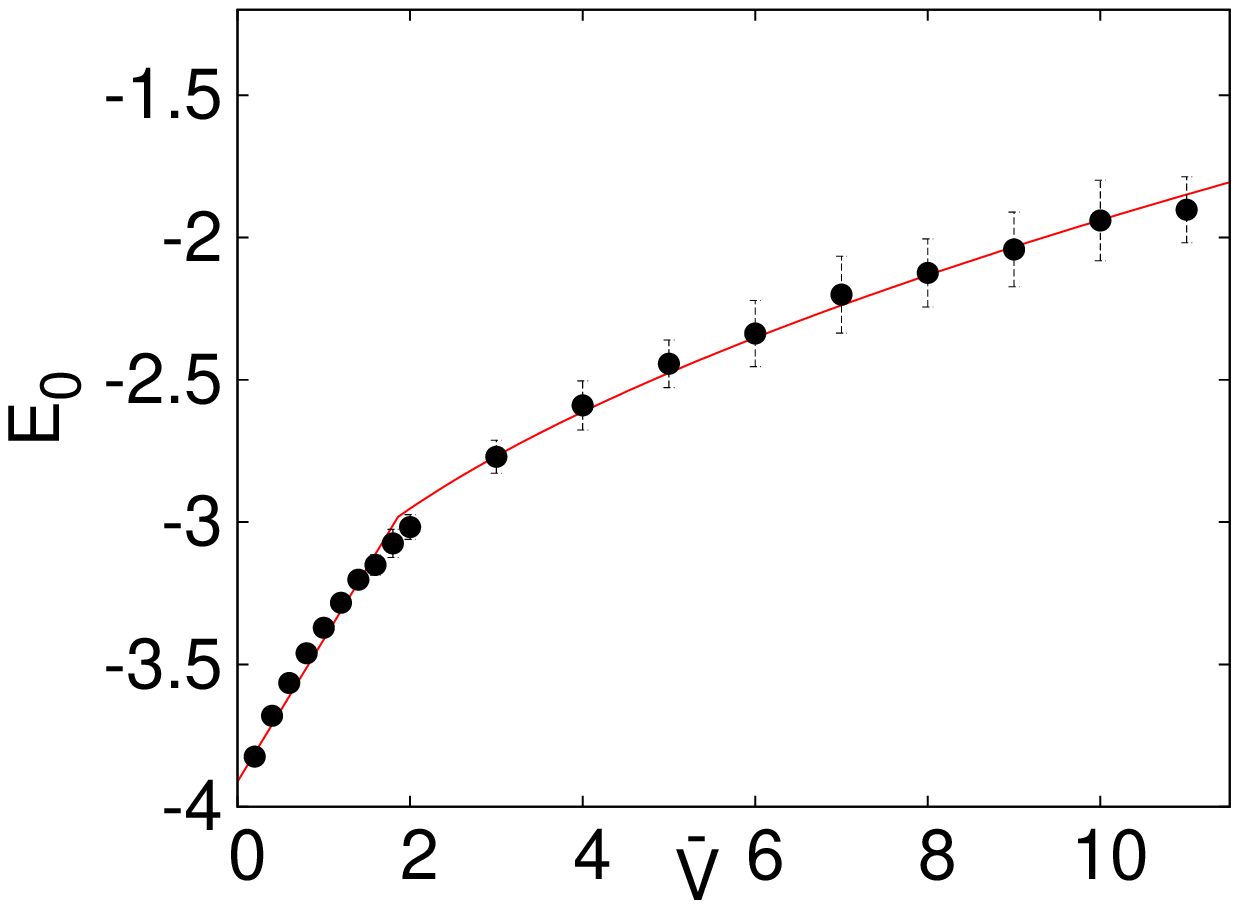}
\caption{}
\end{subfigure}
\caption{(a) Single particle ground state energy for speckle disorder correlation length $l_c=5$. The black dots are calculated results and red line is the fit function. The fit function is $y=0.25873x^{\frac{1}{2}}-4.00983$. (b) Single particle ground state energy for speckle disorder correlation length $l_c=1$. The fit function is: quantum region $y=0.5x-3.91184$, classical region $y=0.579817x^{\frac{1}{2}}-3.77247$.}
\label{single}
\end{figure}

For single particle problem, the ground state energy is calculated with exact diagonalization method. The results are shown in Fig.~\ref{single}.
For correlation length $l_{c}=5$, the system is in classical region for all the calculated disorder strength. As shown in Fig. \ref{single} (a), the single particle ground state energy is well characterized by function $y=0.25873x^{\frac{1}{2}}-4.00983$. The 0.5 power is expected from the trapping scenario since in classical region $\bar{V}\gg E_c$, the wave length of single particle is smaller than disorder correlation length and therefore can be characterized by single fermion in a harmonic trap. The harmonic trap is characterized by average disorder strength and correlation length of speckle disorder potential after disorder average. This is also the case for the classical region for $l_c=1$. As shown in Fig. \ref{single} (b), when average disorder strength $\bar{V}$ is larger than 2.0, the system is in the classical region and the ground state energy is well characterized by the 0.5 power fit function $y=0.579817x^{\frac{1}{2}}-3.77247$. 

The more interesting result is the ground state energy in the quantum region when average disorder strength $\bar{V}$ is smaller than the correlation energy and thus can't be explained by the simple trapping scenario as shown in Fig. \ref{single} (b). This in fact corresponds to the Anderson localization scenario with large localization length. This can be confirmed from the following arguments. 

\begin{figure}[!t]
\centering
\includegraphics[scale=0.45]{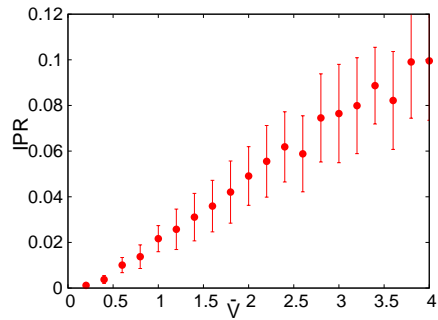}
\caption{Inverse participation ratio (IPR) for single particle under speckle disorder potential with correlation length $l_c=1$.}
\label{singleIPR}
\end{figure}

First, one could calculate the inverse participation ratio for single particle under speckle disorder potential for correlation length $l_c=1$. The result is shown in Fig. \ref{singleIPR}.
Compared with Harper model results (see Fig. \ref{phase1}), the inverse participation ratio is always finite even in the quantum region where disorder strength is smaller than correlation energy and it increases asymptotically with increasing disorder strength and there is no abrupt change. This result implies that there is no phase transition from delocalized phase to localized phase for the case of speckle disorder potential. The particle is in the localized state with large localization length when disorder strength is small and the localization length decreases when disorder strength increases.  

The other rough argument is that we can use the localized wave function as the single particle ground state wave function to calculate the ground state energy. The normalized wave function is:
\begin{equation}
\psi(r)=\frac{1}{\zeta}\sqrt{\frac{2}{\pi}}\exp{\left(\frac{-r}{{\zeta}}\right)},\label{singlewave}
\end{equation}
here, $\zeta$ is the localization length.

The Schr\"odinger equation in two dimensions is:
\begin{equation}
-\frac{1}{2}\left[\frac{\partial^2}{\partial r^2}+\frac{1}{r}\frac{\partial}{\partial r}+\frac{1}{r^2}\frac{\partial^2}{\partial \theta^2}\right]\psi+V(r)\psi=E\psi.
\end{equation}
$V(r)$ is the speckle disorder potential. Using Eq. \eqref{singlewave}
and multiply $\psi^*(r)$ on the left and integrate over the two dimensional space, one could obtain:
\begin{equation}
\frac{1}{2\zeta^2}+\frac{4\pi}{\pi\zeta^2}\int_{0}^{\infty}V(r)\exp{\frac{-2r}{\zeta}}rdr=E.
\end{equation}
When localization length is large, we could approximate $\exp{\frac{-2r}{\zeta}}=1$ and integrate till $r=\frac{\zeta}{2}$ where $\exp{\frac{-2r}{\zeta}}$ reaches $1/e$ of its maximum value (similar to the way correlation length $l_c$ of speckle patten is defined in experiments \cite{bosons,fermions}), then the equation becomes:
\begin{equation}
\begin{aligned}
&\frac{1}{2\zeta^2}+\frac{4\pi}{\pi\zeta^2}\int_{0}^{\frac{\zeta}{2}}V(r)rdr=E, \\
&\frac{1}{2\zeta^2}+\frac{1}{2}\frac{2\pi\int_{0}^{\frac{\zeta}{2}}V(r)rdr}{\pi({\frac{\zeta}{2}})^2}=E.
\end{aligned}
\end{equation}
When localization length is large compared with speckle disorder correlation length, 
$\frac{2\pi\int_{0}^{\frac{\zeta}{2}}V(r)rdr}{\pi({\frac{\zeta}{2}})^2}$ can be regarded as the average value of disorder strength. Then,
\begin{equation}
\frac{1}{2\zeta^2}+\frac{1}{2}\bar{V}=E,
\end{equation}
here, $\bar{V}$ is the average value of disorder strength. This agrees with the computational result in the quantum region (see Fig. \ref{single} (b)). 

\begin{figure}[H]
\centering
\begin{subfigure}[b]{0.2\textwidth}
\includegraphics[width=3.6cm]{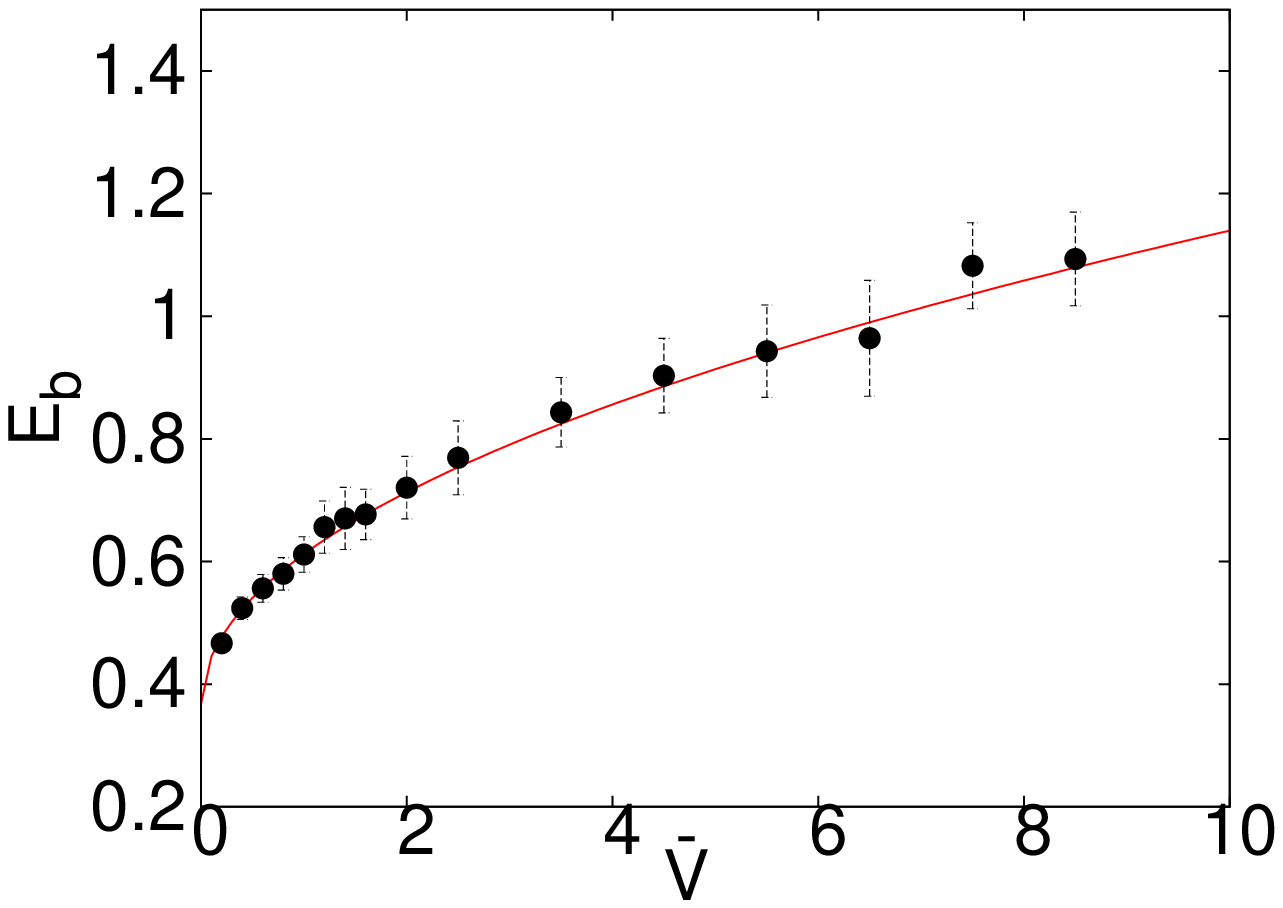}
\caption{}
\end{subfigure}
\begin{subfigure}[b]{0.2\textwidth}
\includegraphics[width=3.6cm]{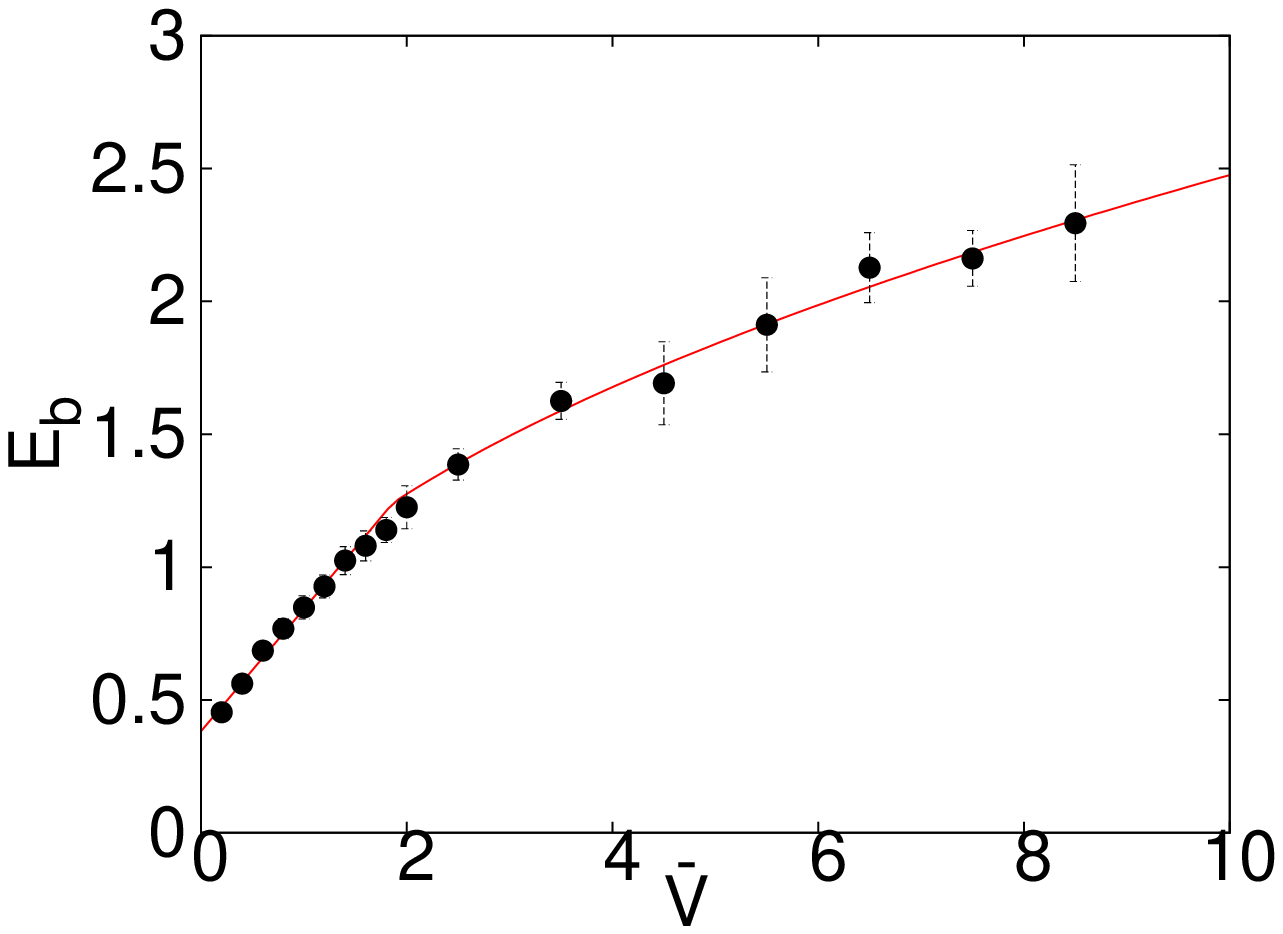}
\caption{}
\end{subfigure}
\begin{subfigure}[b]{0.2\textwidth}
\includegraphics[width=3.6cm]{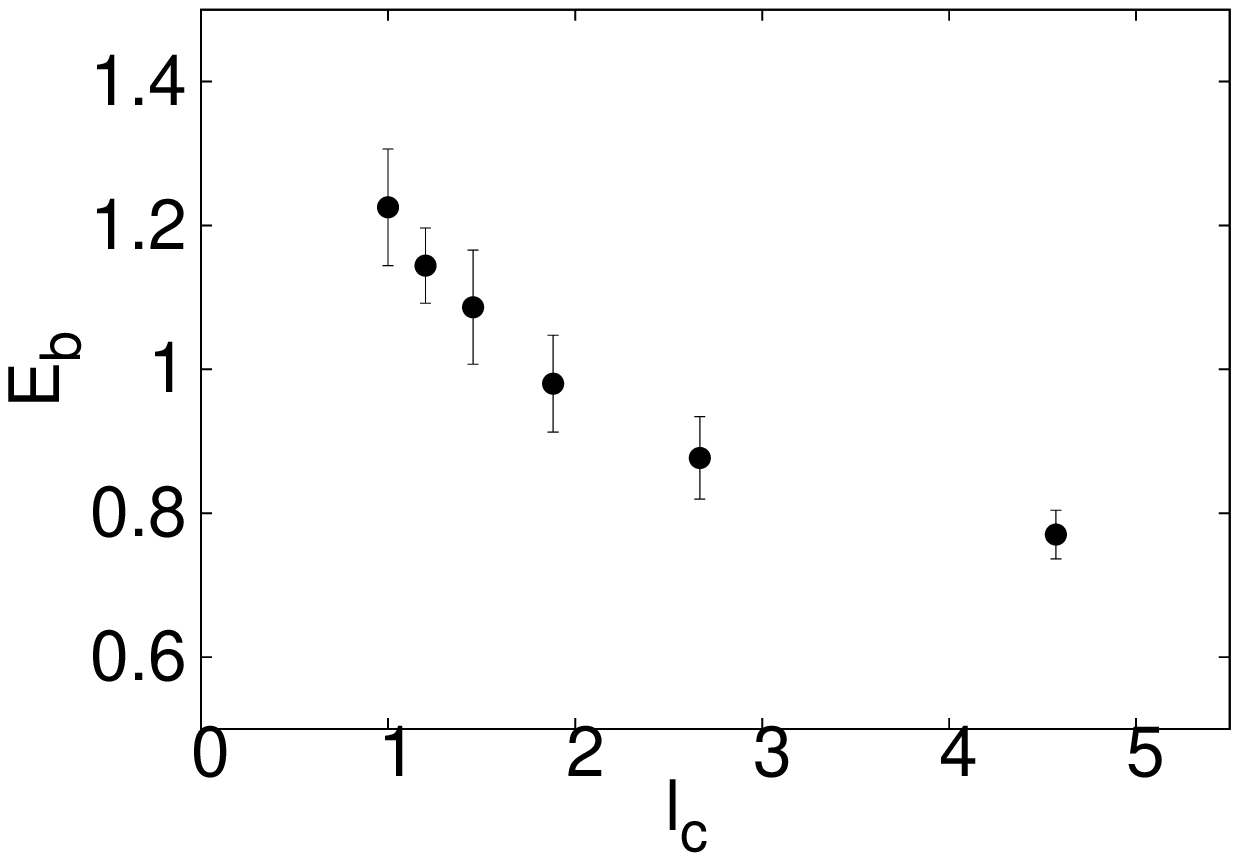}
\caption{}
\end{subfigure}
\begin{subfigure}[b]{0.2\textwidth}
\includegraphics[width=3.6cm]{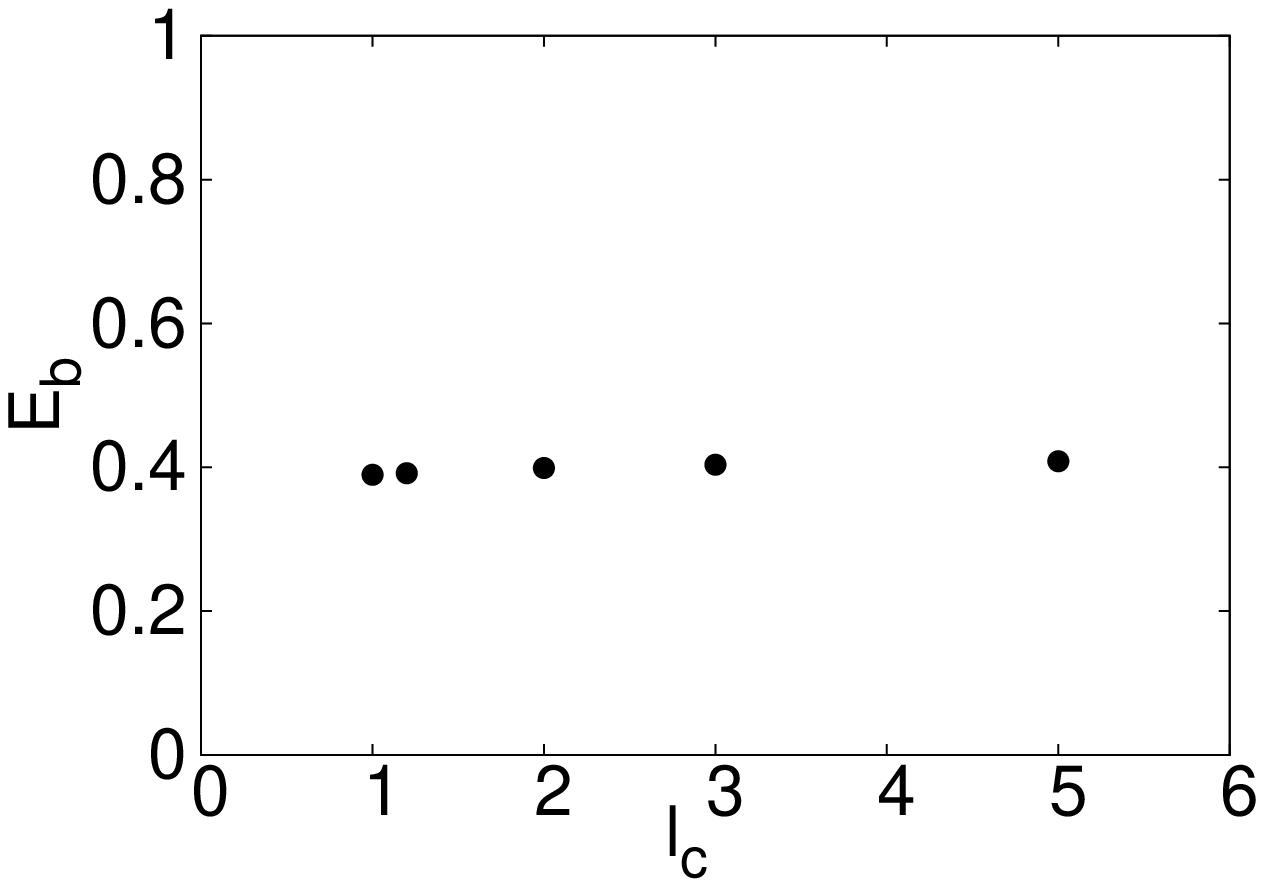}
\caption{}
\end{subfigure}
\caption{(a) Binding energy for increasing disorder strength with correlation length $l_c=5$. The black dots are calculated results and the red line is the fit function. The fit function is  $y=0.244014x^{\frac{1}{2}}+0.368016$. (b) Binding energy for increasing disorder strength with correlation length $l_c=1$. Fit function: quantum region $y=0.46034x+0.3821$, classical region $y=0.686641x^{\frac{1}{2}}+0.304171$. (c) Binding energy for different disorder correlation length $l_c$. Disorder strength $\bar{V}=2.0$ and all data points are in classical region. (d) Disorder strength $\bar{V}=0.05$, all data points are in quantum region.}
\label{twoparticlespeckle}
\end{figure}

Now, we consider two fermions (one spin up and one spin down) with large on-site attractive interaction under speckle disorder potential. The attractive interaction strength is fixed at $|U|=5t$ and we focus on their binding energy. The definition of binding energy is still $E_{b}=2E_{0}-E$. First, we calculate the binding energy for the case of correlation length $l_c=5$ and all calculated data points are in classical region. The results are shown in Fig. \ref{twoparticlespeckle} (a).

The binding energy result can be well approximated with $1/2$ power function over disorder strength. This agrees with the two-particle binding energy result in a harmonic trap \cite{fermions,twoatominharmonictrap}. Another evidence for the trapping scenario is the binding energy for different correlation length in the classical region. Here, after disorder average, the trap is characterized by average disorder strength $\bar{V}$ and correlation length $l_c$. Increasing correlation length has the effect of decreasing the frequency of the trap and hence decreases the two-particle binding energy as shown in Fig. \ref{twoparticlespeckle} (c) \cite{twoatominharmonictrap}. 

\begin{figure}[!t]
\centering
\begin{subfigure}[b]{0.24\textwidth}
\includegraphics[width=4.2cm]{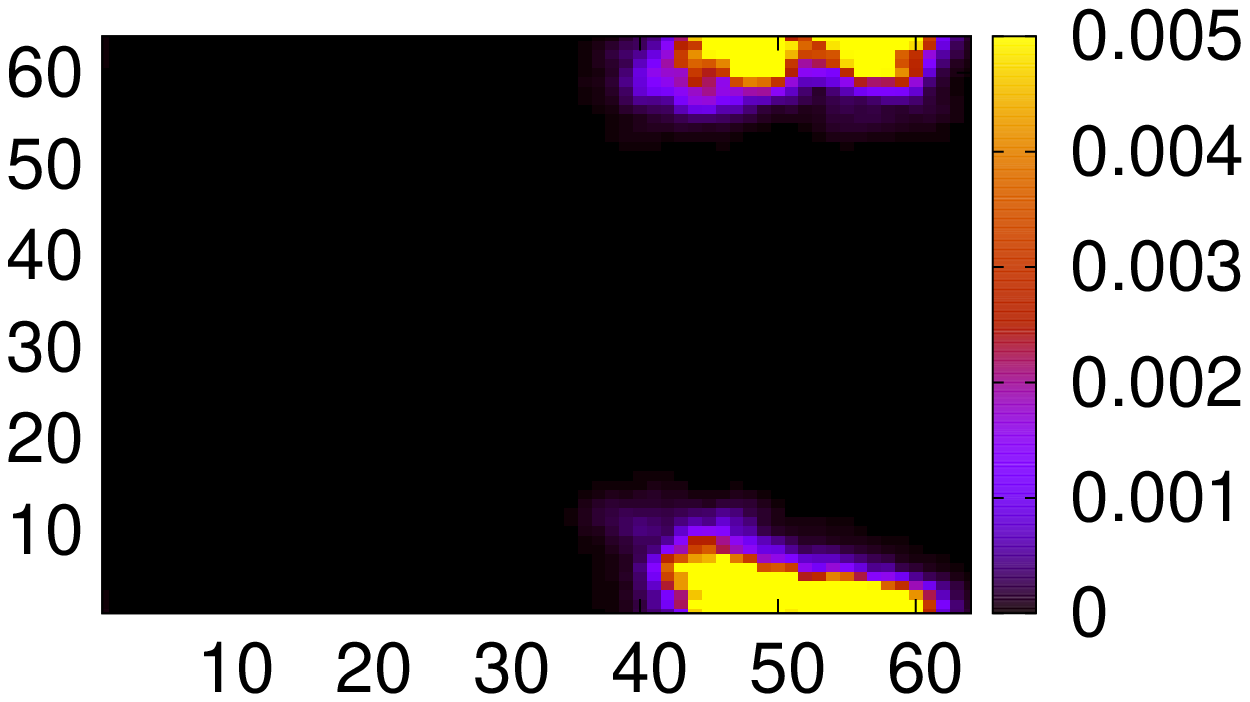}
\caption{}
\end{subfigure}
\begin{subfigure}[b]{0.2\textwidth}
\includegraphics[width=4.2cm]{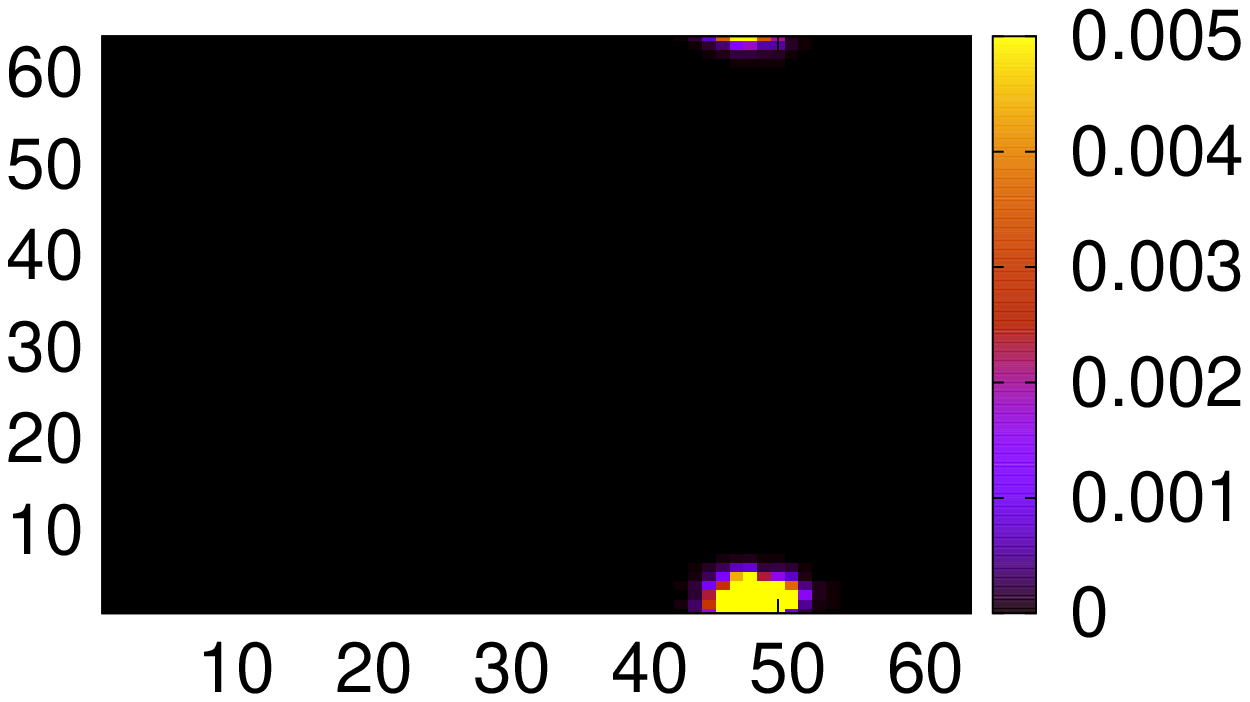}
\caption{}
\end{subfigure}
\begin{subfigure}[b]{0.24\textwidth}
\includegraphics[width=4.2cm]{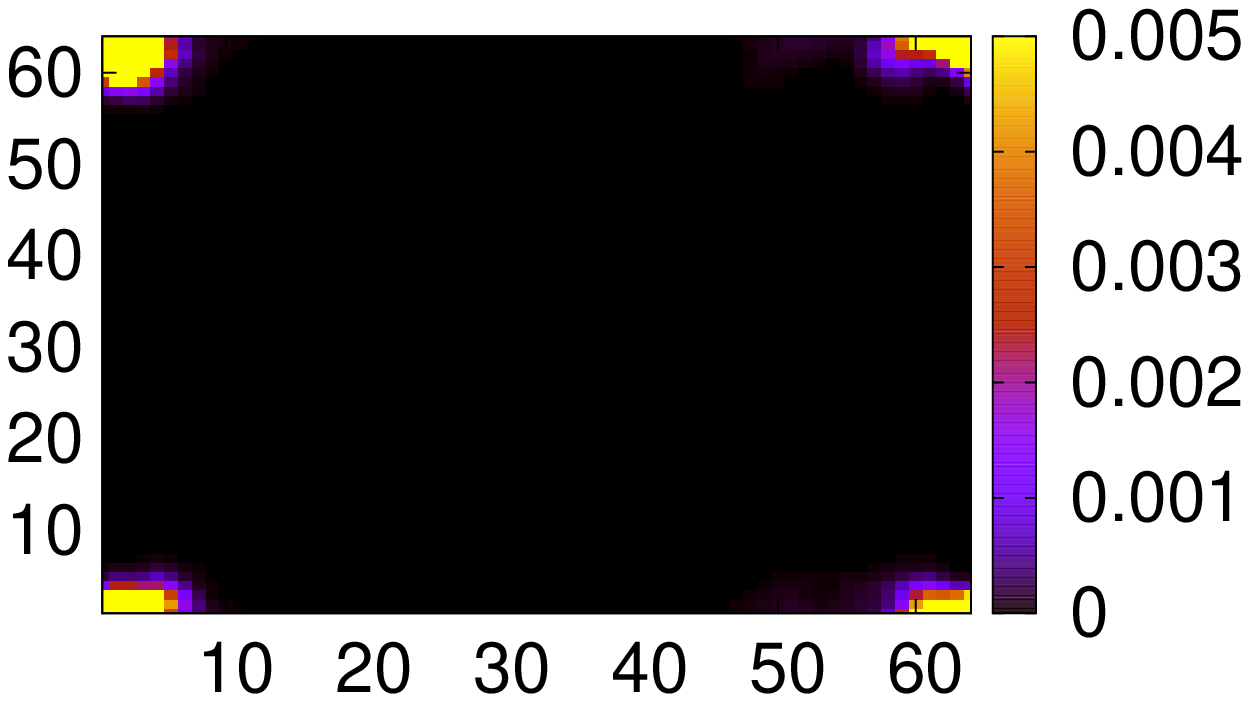}
\caption{}
\end{subfigure}
\begin{subfigure}[b]{0.2\textwidth}
\includegraphics[width=4.2cm]{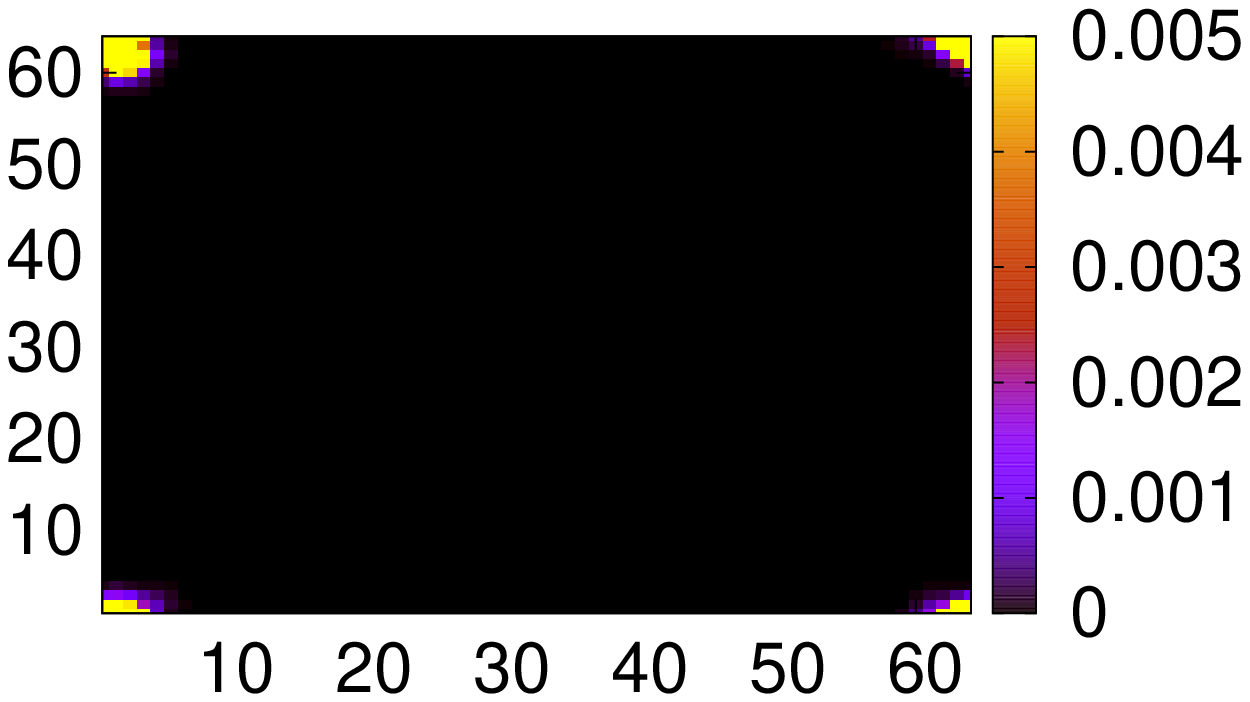}
\caption{}
\end{subfigure}
\caption{(Color online) Local density distribution for speckle pattern with (a) correlation length $l_c=1$, average disorder strength $\bar{V}=0.2$, (b) correlation length $l_c=1$, average disorder strength $\bar{V}=1.0$, (c) correlation length $l_c=5$, average disorder strength $\bar{V}=0.2$, (d) correlation length $l_c=5$, average disorder strength $\bar{V}=1.0$. (a) and  (b) are under the same speckle disorder configuration, (c) and (d) are under the same speckle disorder configuration. Periodic boundary condition is used in the calculation.}
\label{localdensitydistribution}
\end{figure}

We also calculate the binding energy for different disorder strength with disorder correlation length $l_c=1$. As shown in the Fig. \ref{twoparticlespeckle} (b), it is obvious that there are two different regions, one is the quantum region ($\bar{V} \ll E_c$) and the other classical region ($\bar{V}\gg E_c$). The binding energy in the classical region can also be well fitted with the $1/2$ power fit function as a result of the trapping scenario as explained earlier. Then, we focus on the quantum region and calculate the effect of correlation length on binding energy. As shown in Fig.~\ref{twoparticlespeckle} (d), the binding energy in quantum region is almost not affected by the correlation length of speckle disorder, which is very different with the result in the classical region as shown in Fig. \ref{twoparticlespeckle} (c). First, this provides the evidence for the validity of percolation theory in the quantum region since the details of the disorder are not important \cite{fermions}. What is more, since the disorder correlation length has no effect on the binding energy, the behavior of system in the quantum region can't be simply explained by the trapping scenario. Since the definition of binding energy is $E_{b}=2E_{0}-E$ and the linear increase of $E_b$ and $E_0$ in the quantum region, one could see the linear increase of two-particle ground state energy. The similar behaviour of two-particle ground state energy and single particle ground state energy (see Fig. \ref{single}) implies that the two-particle ground state can be characterized by a state similar to the single particle localized state and the localization length decreases when disorder strength increases. This can also be shown from the local density distribution directly. Local density distribution is defined as \cite{andersonlocalizationofpair}:
\begin{equation}
n_i=2\sum_m|\psi(i,m)|^2,
\end{equation}
$\psi(i,m)$ is the two-particle ground state wave function. Factor of 2 comes from the particle number. Some typical results of local density distribution are shown in Fig. \ref{localdensitydistribution}. 

As shown in Fig. \ref{localdensitydistribution}, it is obvious that the wave function is localized although the localization length is large compared with disorder correlation length when the disorder strength is small (see Fig. \ref{localdensitydistribution} (a) and (c)). When the disorder strength increases, the localization length decreases (see Fig. \ref{localdensitydistribution} (b) and (d)). This is in agreement with the single particle case as discussed earlier that there is no phase transition from delocalized phase to localized phase. This can also be understood since the strongly interacting two particles can be described by the effective single particle as discussed earlier for the case of two interacting particles under quasi-periodic potential. 

More convincing results would be the inverse participation ratio calculation after disorder average, but this is too time consuming for our limited computational capacity especially for large lattice size which is required by the statistical properties of speckle disorder potential. Then we resort to the calculation of transport property. Motivated by the recent experiments on transport property \cite{conduction,RC,bosons,fermions}, we focus on Drude weight for increasing attractive interaction strength with speckle disorder potential in the quantum region. Drude weight provides a method to explore transport property of disordered system and it can be calculated from the dependence of the ground state energy $E_0$ on $\phi$ which is the
flux through a torus ring \cite{Kohn,twist}:
\begin{equation}
\frac{D}{\pi}=N\left[\frac{\partial ^{2} E_{0}(\phi)}{\partial {\phi} ^{2}}\right]_{\phi=0},
\end{equation}
in the limit of a large system, D would decrease exponentially to zero for an insulating state while remain finite for a metallic state. In the numerical calculation, one needs to be careful to follow the ground state adiabatically because there is level crossing between ground state and excited state when the twist boundary condition is used \cite{scz1,scz2}. This method has already been used in one-dimensional and two-dimensional system with random disorder potential and repulsive interaction \cite{onedrudeweight,twodmetalinsulator}. We adopt this method to study the transport property under speckle disorder potential with attractive interaction.

\begin{figure}[!t]
\centering
\includegraphics[scale=0.4]{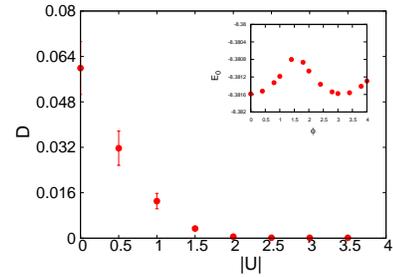}
\caption{Drude weight for different interaction strength under speckle disorder potential. The disorder strength is $\bar{V}=0.5$ and correlation length $l_c=1$. The insert figure shows ground state energy for different flux as illustration.}
\label{drudeweight}
\end{figure}

Fig. \ref{drudeweight} gives the Drude weight result for different attractive interaction strength. The disorder strength is chosen at $\bar{V}=0.5$ and correlation length is $l_c=1$ so it is in the quantum region. As shown in the result, when interaction is large, Drude weight becomes vanishingly small which indicates that the system is in the localized state. This is in agreement with our earlier analysis. So, with all the arguments from the two-particle binding energy, local density distribution and Drude weight, one can conclude that the picture is that the strongly interacting system starts with a localized state in the quantum region which has large localization length compared with disorder correlation length and evolves to a localized state which can be well characterized by the trapping scenario in the classical region. From this point of view, it is not a phase transition from delocalized state to localized state which is different with the two-dimensional Harper model. But as for the real many-body system, from the percolation view, it could result in a percolation transition because the localization length for each pair of fermions changes from large localization length to small localization length compared with the correlation length of speckle disorder potential when disorder strength increases and this could result in a percolation transition. This is helpful to understand the experimental phenomena revealed in the experiment~\cite{fermions}. In this experiment, the transport property of strongly interacting fermions can be well fitted with percolation simulation result and the percolation transition happens when the system changes from the quantum region to classical region as the disorder strength increases. It is worth noting that the disorder correlation length in the experiment is about one pixel in their percolation analysis which is equal to one lattice size in our simulation. As shown in earlier analysis, for one pair, the localization length decreases from a value larger than disorder correlation length to a value smaller than disorder correlation length when the system changes from quantum region to classical region. This could corresponds to the percolation transition in the experiment.  

On the other hand, for small attractive interaction strength, the interaction seems to decrease the Drude weight and make the system more localized than the non-interacting case. This is different from the results and analysis in the experiment \cite{2d quasi experiment}. It is expected that the small interaction would be helpful for the thermalization especially when the initial state is localized state. The contradiction probably comes from the fact that the simulation results only applies to a two-particle system and is inadequate for the description of interaction effect in the real many-body interacting system. 

\section{\label{sec:level1} IV. summary}
The one- and two-particle problem with quasi-periodic and speckle disorder potential is examined using exact diagonalization (ED) method. The phase diagram for two-dimensional quasi-periodic disorder potential is given and the effect of strong interaction is also discussed. For speckle pattern, the different behaviour of single particle ground state energy and two-particle binding energy in the quantum region and classical region is discussed. We also calculate the Drude weight for different interaction in order to show the effect of interaction for the speckle disorder potential case.

\section{\label{sec:level1}acknowledgements}
The author would like to thank Zhihao Xu and Shizhong Zhang for useful discussions and critical reading of the manuscript. This work is supported by Hong Kong Research
Grants Council (General Research Fund, HKU 17318316 and Colaborative Research Fund, C6026-16W).


\begin{thebibliography}{20}
\bibitem{50anderson}Edited by Elihu Abrahams. \textit{50 Years of Anderson Localization}. World Scientific, 2010.
\bibitem{conductorinsulator} V. Dobrosavljevic, N. Trivedi and J. M. Valles, \textit{Conductor-insulator quantum phase transition}. Oxford, 2012.
\bibitem{mesoscopic1}Ping Sheng, \textit{Introduction to wave scattering, localization, and mesoscopic phenomena}. Academic,
London, 1995.
\bibitem{mesoscopic2}Supriyo Datta, \textit{Electronic transport in mesoscopic systems}, Cambridge University Press 1995.
\bibitem{anderson}P. W. Anderson, Phys. Rev. {\bf 109}, 1492 (1958).
\bibitem{scaling} E. Abrahams, P. W. Anderson, D. C. Licciardello, and T. V. Ramakrishnan, Phys. Rev. Lett. {\bf 42}, 673, (1979).
\bibitem{lightlocalization1} D. S. Wiersma, P. Bartolini, A. Lagendijk and Roberto Righini, Nature, {\bf 390}, 671 (1997).
\bibitem{lightlocalization2} T. Schwartz, G. Bartal, S. Fishman and M. Segev, Nature, {\bf 446}, 52, (2007).
\bibitem{microwavelocalization} R. Dalichaouch, J. P. Armstrong, S. Schultz, P. M. Platzman and S. L. McCall. Nature, {\bf 354}, 53 (1991). 
\bibitem{soundlocalization} R. L. Weaver, Wave Motion, {\bf 12}, 129 (1990). 
\bibitem{1dal1}J. Billy, V. Josse, Z. Zuo, A. Bernard, B. Hambrecht, P. Lugan, D. Cl ́ement, L. Sanchez-Palencia, P. Bouyer and A. Aspect, Nature {\bf 453}, 891 (2008).
\bibitem{1dal2}G. Roati, C. D’Errico, L. Fallani, M. Fattori, C. Fort, M. Zaccanti, G. Modugno, M. Modugno, M. Inguscio, Nature {\bf 453}, 895 (2008).
\bibitem{3dal1} S. S. Kondov, W. R. McGehee, J. J. Zirbel and B. DeMarco, Science, {\bf 334}, 66 (2011).
\bibitem{3dal2} F. Jendrzejewski, A. Bernard, K. M ̈uller, P. Cheinet, V. Josse, M. Piraud, L. Pezz ́e, L. Sanchez-Palencia, A. Aspect and P. Bouyer, Nat. Phys. {\bf 8}, 398 (2012).
\bibitem{diagonaloffdiagonal}J. C. Flores, J. Phys. Condens. Matter {\bf 1}, 8471 (1989).
\bibitem{RDM1}D. H. Dunlap, H.-L. Wu, and P. W. Phillips, Phys. Rev. Lett. {\bf 65}, 88 (1990).
\bibitem{RDM2} P. Phillips and H.-L. Wu, Science {\bf 252}, 1805 (1991).
\bibitem{singlespeckle}P. Lugan, A. Aspect, L. Sanchez-Palencia, D. Delande, B. Gr\'emaud, C. 
A. M\"uller, and C. Miniatura, Phys. Rev. A. {\bf 80}, 023605 (2009).
\bibitem{1d interaction AAH} M. Schreiber, S. S. Hodgman, P. Bordia, H. P. L\"uschen, M. H. Ficher, R. Vosk, E. Altman, U. Schneider, and I. Bloch, Science {\bf 349}, 842 (2015).
\bibitem{isingexperiment} J. Smith, A. Lee, P. Richerme, B. Neyenhuis, P. W. Hess, P. Hauke, M. Heyl, D. A. Huse, and C. Monroe, Nat. Phys. {\bf 12}, 907, (2016).
\bibitem{quasi2d interacting AAH} P. Bordia, H. P. L\"uschen, S. S. Hodgman, M. Schreiber, I. Bloch, and U. Schneider, Phys. Rev. Lett. {\bf 116}, 140401 (2016).
\bibitem{2d random disorder experiment} J.-y. Choi, S. Hild, J. Zeiher, P. Schau{\ss}, A. Rubio-Abadal, T. Yefsah, V. Khemani, D. A. Huse, I. Bloch, and C. Gross, Science {\bf 352}, 1547 (2016).
\bibitem{2d quasi experiment} P. Bordia, H. L\"uschen, S. Scherg, S. Gopalakrishnan, M. Knap, U. Schneider, and I. Bloch. Phys. Rev. X {\bf 7}, 041047 (2017).
\bibitem{conduction}{J. -P. Brantut, J. Meineke, D. Stadler, S. Krinner, and T. Esslinger. Science {\bf 337}, 1069 (2012)}
\bibitem{RC} D. Stadler, S. Krinner, J. Meineke, J. -P. Brantut, and T. Esslinger. Nature. {\bf 491}, 736 (2012).
\bibitem{bosons} S. Krinner, D. Stadler, J. Meineke, J. -P. Brantut and T. Esslinger. Phys. Rev. Lett. {\bf 110}, 100601 (2013).
\bibitem{fermions} S. Krinner, D. Stadler, J. Meineke, J. -P. Brantut and T. Esslinger. Phys. Rev. Lett. {\bf 115}, 045302 (2015).
\bibitem{harper}P. G. Harper, Proc. Phys. Soc. Lond. A {\bf 68} 874 (1955). 
\bibitem{aubry andre} Aubry and Andr{\'e} G, Ann. Isr. Phys. Soc. {\bf 3} 33, 1980.
\bibitem{specklepattern} J. W. Goodman. \textit{Speckle phenomena in optics: theory and applications}. Chapter 2. Roberts and Company Publishers, 2007.
\bibitem{simulationspeckle} J. M. Huntley. Applied optics, {\bf 28}, 20 (1989).
\bibitem{computationalmanyparticle} Edited by H. Fehske, R. Schneider, A. Wei{\ss}e, \textit{Computational many-particle physics}, Chapter 18. Springer 2008.
\bibitem{tworandom}D. L. Shepelyansky, Phys. Rev. Lett. {\bf 73}, 2607 (1994). 
\bibitem{imry} Y. Imry, Europhys. Lett. {\bf 30}, 405 (1995).
\bibitem{frahm} K. Frahm, A. M\"uller-Groeling, J.-L. Pichard, and D. Weinmann. Europhys. Lett. {\bf 31}, 169 (1995). 
\bibitem{weinmann}D. Weinmann, A. M\"uller-Groeling, J.-L. Pichard, and K. Frahm. Phys. Rev. Lett. {\bf 75}, 1598 (1995).
\bibitem{oppen} F. von Oppen, T. Wettig, and J. M\"uller, Phys. Rev. Lett. {\bf 76}, 491 (1996).
\bibitem{twoharper}D. L. Shepelyansky, Phys. Rev. B {\bf 54}, 14896 (1996).
\bibitem{ueffect}B. Georgeot and D. L. Shepelyansky, Phys. Rev. Lett. {\bf 79}, 4365 (1997). 
\bibitem{andersonlocalizationofpair} G. Dufour and G. Orso, Phys. Rev. Lett, {\bf 109},155306 (2012).
\bibitem{bosonspeckle}Pere Mujal, Artur Polls, Sebastiano Pilati, and Bruno Juli\'a-D\'iaz, Phys. Rev. A, {\bf 100}, 013603 (2019).
\bibitem{onedrudeweight}M. Filippone, P. W. Brouwer, J. Eisert, and F. von Oppen, Phys. Rev. B. {\bf 94}, 201112 (2016).
\bibitem{1d boson} S. Flach, M. Ivanchenko, and R. Khomeriki, European
Phys. Lett. {\bf 98}, 66002 (2012).
\bibitem{quantumclassicregion} B. I. Shklovskii, Semiconductors {\bf 42}, 909-913 (2008).
\bibitem{twoatominharmonictrap} T. Bushch, B.-G. Englert, K. Rzazewski, and M. Wilkens, Foundations of Physics {\bf 28}, 549 (1998).
\bibitem{Kohn} W. Kohn, Phys. Rev. {\bf 133}, A171 (1964).
\bibitem{twist} B. S. Shastry and B. Sutherland, Phys. Rev. Lett. {\bf 65}, 243 (1990).
\bibitem{scz1} D. J. Scalapino, S. R. White, S.-C Zhang, Phys. Rev. Lett. {\bf 68}, 2830 (1992). 
\bibitem{scz2} D. J. Scalapino, S. R. White, S.-C Zhang, Phys. Rev. B, {\bf 47}, 7995 (1993).
\bibitem{twodmetalinsulator} R. Kotlyar and S. Das Sarma, Phys. Rev. Let, {\bf 86}, 2388 (2001).

\end{thebibliography}
\end{document}